\documentclass[]{raa}
\usepackage{graphicx,times}
\usepackage{natbib}
\usepackage{amssymb,amsmath}
\bibpunct{(}{)}{;}{a}{}{,}

\usepackage[a4paper=true,dvipdfm=true,pagebackref=true]{hyperref}
\hypersetup{pdftitle = DMFT IRV YSO, pdfauthor = Khaibrakhmanov, pdfsubject= The subject, pdfkeywords = keyword1 keyword2 keyword3} 
\hypersetup{colorlinks = true, linkcolor = green, anchorcolor = red, citecolor = blue, filecolor = red, pagecolor = red, urlcolor = red}

\begin{document}
   \title{Dynamics of magnetic flux tubes and IR-variability of \\
   Young Stellar Objects
$^*$
\footnotetext{\small $*$ Supported by the Russian science foundation.}
}

 \volnopage{ {\bf 2017} Vol.\ {\bf X} No. {\bf XX}, 000--000}
   \setcounter{page}{1}

   \author{Khaibrakhmanov Sergey\inst{1,2}, Dudorov Alexander\inst{2}, Sobolev Andrey  \inst{1}
   }

   \institute{Kourovka astronomical observatory, Ural Federal University, Ekaterinburg 620000, 
Russia; {\it khaibrakhmanov@csu.ru}\\
        \and
             Theoretical physics department, Chelyabinsk state university,
             Chelyabinsk 454001, Russia\\
\vs \no
   {\small Received 2017 November 30; accepted 2017 December 25}
}

\abstract{We simulate the dynamics of slender magnetic flux tubes (MFTs) in the accretion disks of T~Tauri stars. The dynamical equations of our model take into account the aerodynamic and turbulent drag forces, and the radiative heat exchange between the MFT and ambient gas. The structure of the disk is calculated with the help of our MHD model of the accretion disks. We consider the MFTs formed at the distances $0.027-0.8$~au from the star with various initial radii and plasma betas $\beta_0$. The simulations show that the MFT with weak magnetic field ($\beta_0=10$) rise slowly with speeds less than the sound speed. The MFTs with $\beta_0=1$ form outflowing magnetized corona of the disk. Strongly magnetized MFTs ($\beta_0=0.1$) can cause the outflows with velocities $20-50$~km s$^{-1}$. The tubes rise periodically over times from several days to several months according to our simulations. We propose that periodically rising MFTs can absorb stellar radiation and contribute to the IR-variability of Young Stellar Objects.
\keywords{accretion disks, instabilities, magnetohydrodynamics (MHD), protoplanetary disks}
}

   \authorrunning{S. Khaibrakhmanov et al. }
   \titlerunning{Magnetic flux tubes dynamics and YSOs variability}
   \maketitle

\section{Introduction}
\label{sect:Intro}

Various observations indicate that stars at the early stages of the evolution (Young Stellar Objects, YSOs~\citep{adams87}) have magnetic field. Measurements of Zeeman splitting of the spectral lines show that classical T Tauri stars (Class II YSOs) have surface magnetic field with strength $1-3$~kG~\citep{krull07}. Polarization maps of the dust thermal emission indicate that the accretion disks of the young stars have large-scale magnetic field~\citep{li16, li18}. The angular resolution is yet not enough to detect the magnetic field geometry in details.

Theory of fossil magnetic field predicts that the magnetic field of the accretion disks of young stars originates from the magnetic field of parent molecular cloud (see for review~\cite{dudorov95, fmft}). The dynamo mechanism driven by the turbulent cyclonic motions and differential rotation in the conducting plasma can also lead to the generation of the magnetic field in the disk (see, for example, reviews \cite{brandenburg05}, \cite{blackman12}). Some aspects of the dynamo action in the accretion disks can be found in  \cite{brandenburg95}, \cite{kitchatinov10},  \cite{gressel15}, \cite{moss16}. In this work, we use the MHD model of the accretion disks of young stars with fossil magnetic field developed by~\cite{fmfadys}. \cite{fmfadys} have shown that the magnetic field geometry varies through the disk. The magnetic field is quasi-azimuthal, $B_{\varphi}\sim B_z$ (in cylindrical coordinates), in the  regions of thermal ionization, where the magnetic field is frozen in gas. Throughout most of the disk Ohmic diffusion, magnetic ambipolar diffusion and the Hall effect operate (for example, see review~\cite{turner14}). The magnetic field is quasi-poloidal, $(B_r,\,B_{\varphi})\ll B_z$, inside the regions of low ionization fraction (``dead'' zones,~\cite{gammie96}), and quasi-azimuthal or quasi-radial, $B_r\sim B_z$, in the outer regions depending on the ionization parameters.  The magnetic field has quasi-radial geometry near the borders of the ``dead'' zones~\citep{kh17}.

The intensity of azimuthal magnetic field $B_{\varphi}$ is amplified over a time scale of order of the rotation period $P_{\rm{orb}}$. The intensity of $B_r$ is amplified over an accretion time scale $t_{\rm{acc}}$. In the accretion disks $P_{\rm{orb}}\ll t_{\rm{acc}}$, therefore $B_r\ll B_{\varphi}$ in the inner region (see \cite{fmfadys}). The problem is what mechanism hinders the significant growth of the azimuthal magnetic field in the region of thermal ionization. \cite{fmfadys} have assumed that the magnetic buoyancy can be such a mechanism. Magnetic flux tubes (MFTs) form as a result of Parker instability of a gas layer with strong horizontal magnetic field (see~\cite{parker_book}). A number of numerical simulations confirmed the development of the Parker instability and MFTs formation~\citep{cattaneo88, matthews95, wissink00, fan01, vasil08}. Once formed, the MFTs rise from the disk under the action of the buoyancy force. This process leads to escape of the excess magnetic flux from the regions of its generation. \cite{kh17} and \cite{kh17b} incorporated the magnetic buoyancy into the induction equation and showed that the buoyancy can be treated as the additional mechanism of the magnetic flux escape from the disks.

Usually, the MFT dynamics in the accretion disks has been investigated in frame of slender flux tube approximation~\citep{sakimoto89, torkelsson93, chakra94, schram96, bmfad}. The dynamics is determined by the buoyancy force, drag forces, thermal structure of the disk, efficiency of the heat exchange with ambient gas, relation between the centrifugal and magnetic tension forces.

\cite{bmfad} considered the dynamics of slender adiabatic MFTs in the accretion disks of young stars. In this paper, we extend their approach by including the radiative heat exchange in the model equations. In frame of the slender flux tube approximation, we investigate the MFTs dynamics in the accretion disks of T~Tauri stars. The initial parameters take into account the disk structure determined using our MHD model of the accretion disks~\citep{fmfadys, kh17}. Apart from the other researchers, we take into account the turbulent drag inside the disk.

The paper is organized as follows. In Section~\ref{Sec:model}, we present our model of the magnetic flux tubes dynamics. The model of the accretion disk is briefly discussed in Section~\ref{Sec:disk}. Section~\ref{Sec:res} presents results of numerical simulations. Typical dynamics of the MFT is considered in Section~\ref{Sec:fiduc}. Dependence on the model parameters is investigated in Section~\ref{Sec:param}. We make analytical estimates of the terminal MFT velocities in Section~\ref{Sec:v_b}. We search for observational appearance of the MFTs dynamics in Section~\ref{Sec:outflows}. Section~\ref{Sec:discussion} summarizes and discusses our findings.

\section{Model of magnetic flux tube dynamics}
\label{Sec:model}

We investigate the MFTs dynamics inside the accretion disks using the cylindrical coordinates $(r,\, \varphi, \,z)$. Axis $z$ coincides with the disk rotation axis. The magnetic field inside the disk has components ${\bf B}=(B_r,\,B_{\varphi},\,B_z)$. We assume that toroidal magnetic field ${\bf B}_{\rm{t}}=(0,\,B_{\varphi},\,0)$ splits into the magnetic flux tubes due to Parker instability~\citep{parker_book}. The MFT has the form of torus around the disk rotation axis with major radius $a_{\rm{maj}}=r$ and minor radius $a\ll a_{\rm{maj}}$. The MFT is azimuthally symmetric. Therefore, we can investigate motion of the small part of the torus, i.e. cylinder of unit length. This cylindrical MFT has radius $a$, gas pressure $P_{\rm{g}}$, density $\rho$, temperature $T$, and magnetic field strength $B_{\varphi} = B$. The accretion disk is characterized by pressure $P_{\rm{e}}$, density $\rho_{\rm{e}}$, and temperature $T_{\rm{e}}$.

We model the dynamics of the MFT following~\cite{dk85} and use the system of equations
\begin{eqnarray}
	\rho\frac{d{\bf v}}{dt} &=& \left(\rho - \rho_{\rm{e}}\right){\bf g} + \rho{\bf f}_d\left({\bf v},\, \rho,\, T,\, a,\, \rho_e\right),\label{Eq:motion}\\
	\frac{d{\bf r}}{dt} &=& {\bf v},\label{Eq:velocity}\\
	M_{\rm{l}} &=& \rho\pi a^2,\label{Eq:mass}\\
	\Phi &=& B\pi a^2,\label{Eq:mflux}\\
	dQ &=& dU + P_{\rm{e}}dV,\label{Eq:dQ}\\
	P_{\rm{g}} + \frac{B^2}{8\pi} &=& P_{\rm{e}},\label{Eq:pbal}\\
	\frac{dP_{\rm{e}}}{dz} &=& -\rho_{\rm{e}}g_z,\label{Eq:Disk}\\
	P_{\rm{g}} &=& \frac{R_{\rm{g}}}{\mu}\rho T,\label{Eq:eos}\\
	U &=& \frac{P_{\rm{g}}}{\rho(\gamma - 1)} + \frac{B^2}{8\pi\rho}.\label{Eq:eos_kalor}
\end{eqnarray}
Equation (\ref{Eq:motion}) is the equation of motion (where ${\bf f}_d$ is the drag force per unit mass), (\ref{Eq:velocity}, \ref{Eq:mass}, \ref{Eq:mflux}) are the definitions of velocity ${\bf v}$, mass $M_{\rm{l}}$ per unit length and magnetic flux $\Phi$ of the MFT, (\ref{Eq:dQ}) is the first law of thermodynamics ($Q$ is the heat per unit mass, $V=1/\rho$ is the specific volume), (\ref{Eq:pbal}) is the balance between internal pressure ($P=P_{\rm{g}} + \frac{B^2}{8\pi}$) and external pressure ($P_{\rm{e}}$), (\ref{Eq:Disk}) is the equation of hydrostatic equilibrium of the disk, (\ref{Eq:eos}) is the equation of state ($R_{\rm{g}}$ is the universal gas constant, $\mu=2.3$ is the molecular weight), (\ref{Eq:eos_kalor}) is the energy per unit mass, $\gamma$ is the adiabatic index.

First term on the right-hand side of Equation~(\ref{Eq:motion})
\begin{equation}
	{\bf F}_{\rm{b}} = \left(\rho - \rho_{\rm{e}}\right){\bf g}
\end{equation}
is the buoyancy force, that is the difference between gravity force and Archimedes force, ${\bf g}$ is the gravitational acceleration.

We study one-dimensional problem of the MFT motion in $z$-direction, then ${\bf v} = (0,\,0,\, v)$, ${\bf r} = (0,\,0,\, z)$, ${\bf g} = (0,\,0,\,g_z)$. Equality (\ref{Eq:pbal}) shows that the MFT is lighter than the ambient gas, i.e. $\rho<\rho_{\rm{e}}$. Therefore, the buoyancy force ${\bf F}_{\rm{b}}=(0,\,0,\,F_{\rm{b}})$ causes the MFT to move upward. Drag force $\rho{\bf f}_{\rm{d}}=(0,\,0,\,\rho f_{\rm{d}})$ counteracts the motion. Aerodynamic drag force (see~\cite{parker_book})
\begin{equation}
	f_{\rm{d}} = -\frac{\rho_{\rm{e}}v^2}{2}\frac{C_{\rm{d}}}{\rho\pi a^2},\label{Eq:fa}
\end{equation}
where $C_{\rm{d}}$ is the drag coefficient $\sim 1$. Turbulent drag force~\citep{pneuman72}
\begin{equation}
f_{\rm{d}} = -\frac{\pi\rho_{\rm{e}}\left(\nu_{\rm{t}}av^3\right)^{1/2}}{\rho\pi a^2},\label{Eq:ft}
\end{equation}
where $\nu_{\rm{t}}$ is the turbulent viscosity. The latter is estimated as~\citep{ss73}
\begin{equation}
	\nu_{\rm{t}} = \alpha v_{\rm{s}} H,\label{Eq:nu_t}
\end{equation}
where $\alpha$ is non-dimensional constant characterizing the turbulence efficiency,
\begin{equation}
	v_{\rm{s}} = \sqrt{\frac{R_{\rm{g}}T_{\rm{e}}}{\mu}}
\end{equation}
is the isothermal sound speed, $H$ is the height scale of the disk.

Turbulent drag force (Eq.~\ref{Eq:ft}) is taken into account inside the disk. Aerodynamics drag force (Eq.~\ref{Eq:fa}) is considered above the disk.

The disk is assumed to be in hydrostatic equilibrium in $z$-direction. Vertical component of stellar gravity,
\begin{equation}
g_z = -z\frac{GM_{\star}}{r^3}\left(1 + \frac{z^2}{r^2}\right)^{-3/2},
\end{equation}
where $M_{\star}$ is the mass of the star.

System of equations (\ref{Eq:motion}-\ref{Eq:eos_kalor}) can be reduced to
\begin{eqnarray}
	\frac{dv}{dt} &=& \left(1 - \frac{\rho_{\rm{e}}}{\rho}\right)g_z + f_d,\label{Eq:v}\\
	\frac{dz}{dt} &=& v.\label{Eq:z}\\
		a &=& a_0\left(\frac{\rho}{\rho_0}\right)^{-1/2},\label{Eq:a}\\
	B &=& B_0\frac{\rho}{\rho_0},\label{Eq:B}\\
		\frac{d\rho}{dt} &=& \frac{h_{\rm{c}}P_T + U_T\rho_e g_z v}{P_T\left(U_\rho - \dfrac{P_{\rm{e}}}{\rho^2}\right) - U_T\left(P_\rho + C_{\rm{m}}\rho\right)},\label{Eq:rho}\\
	\frac{dT}{dt} &=& \frac{\rho_e g_z v\left(U_\rho - \dfrac{P_{\rm{e}}}{\rho^2}\right) + h_{\rm{c}}\left(P_\rho + C_{\rm{m}}\rho\right)}{U_T\left(P_\rho + C_{\rm{m}}\rho\right) - P_T\left(U_\rho - \dfrac{P_{\rm{e}}}{\rho^2}\right)}\label{Eq:T}\\
	\rho_{\rm{e}} &=& \rho_{\rm{m}}e^{-\frac{z^2}{2H}},\label{Eq:rhoe}
\end{eqnarray}
where $a_0$, $\rho_0$ and $B_0$ are the initial radius, density and magnetic field strength of the MFT, $\left(...\right)_T$ means derivative with respect to $T$ (with constant $\rho$), $\left(...\right)_{\rho}$ means derivative with respect to $\rho$ (with constant $T$), $h_{\rm{c}}$ is the heating power per unit mass, $C_{\rm{m}}=\dfrac{B_0^2}{4\pi\rho_0^2}$, $H=v_{\rm{s}}/\Omega$ is the height scale of the disk, $v_{\rm{s}}=\sqrt{R_{\rm{g}}T_{\rm{e}}/\mu}$ is the isothermal sound speed,
\begin{equation}
	\Omega = \sqrt{\frac{GM_{\star}}{r^3}}
\end{equation}
is the keplerian angular velocity.

Equations~(\ref{Eq:a}) and (\ref{Eq:B}) follow from mass and magnetic flux conservation (Eqs. (\ref{Eq:mass}) and (\ref{Eq:mflux})). Equations (\ref{Eq:rho}) and (\ref{Eq:T}) are derived from (\ref{Eq:dQ}, \ref{Eq:pbal}, \ref{Eq:Disk}). Heating power per unit mass is defined as $h_{\rm{c}}=dQ/dt$. In diffusional approximation
\begin{equation}
	h_{\rm{c}} \simeq -\frac{8}{3\kappa_{\rm{R}}\rho^2}\frac{\sigma_{\rm{R}}T^4 - \sigma_{\rm{R}}T_{\rm{e}}^4}{a^2},
\end{equation}
where $\kappa_{\rm{R}}$ is the Rosseland mean opacity, $\sigma_{\rm{R}}$ is the Stefan-Boltzmann constant. We determine $\kappa_{\rm{R}}$ as the power-law function of gas density and temperature following \citet{fmfadys}. 

Formula (\ref{Eq:rhoe}) is the solution of hydrostatic equilibrium Equation (\ref{Eq:Disk}) in the isothermal case, $T_{\rm{e}}=const$. We determine the surface of the disk as the locus $z=3\,H$. Above the surface, temperature is constant $T_{\rm{e}}$ and density falls down with height according to Equation (\ref{Eq:rhoe}) to the point where $\rho_{\rm{e}}$ becomes equal to the density of the interstellar medium $\rho_{\rm{ism}}=3.8\times 10^{-20}\,\rm{g}\,\rm{cm}^{-3}$. 

\section{Model of the disk}
\label{Sec:disk}
We investigate the dynamics of the MFT in the accretion disk of T~Tauri star. We use our MHD model of the accretion disks \citep{fmfadys} to calculate the structure and the magnetic field of the disks. Let us describe briefly the features of the model (see for details \cite{fmfadys} and \cite{kh17}).

The model is MHD-generalization of \cite{ss73} model. We solve the MHD equations in the approximation of a thin stationary disk. It is assumed that the turbulence is the main mechanism of the angular momentum transport. The turbulent viscosity is estimated according to (\ref{Eq:nu_t}). The model has two main parameters: $\alpha$ and accretion rate $\dot{M}$.

The temperature of the disk is calculated from the balance between viscous heating and radiative cooling. We use low-temperature opacities from \cite{semenov03}. The heating by stellar radiation and cosmic rays in the outer parts of the disk are also taken into account.

The magnetic field components are calculated from the induction equation taking into account Ohmic diffusion, magnetic ambipolar diffusion, magnetic buoyancy and the Hall effect. Ionization fraction is determined from the equation of collisional ionization (see~\cite{spitzer_book}) considering the ionization by cosmic rays, X-rays and radioactive decay, radiative recombinations and the recombinations on the dust grains. Additionally, the evaporation of the dust grains and thermal ionization are included in the model.

Outer boundary of the disk, $r_{\rm{out}}$, is determined as the contact boundary, where the disk pressure equals the pressure of the external medium.

\begin{figure}[t]
\centering
	\includegraphics[width=14.0cm, angle=0]{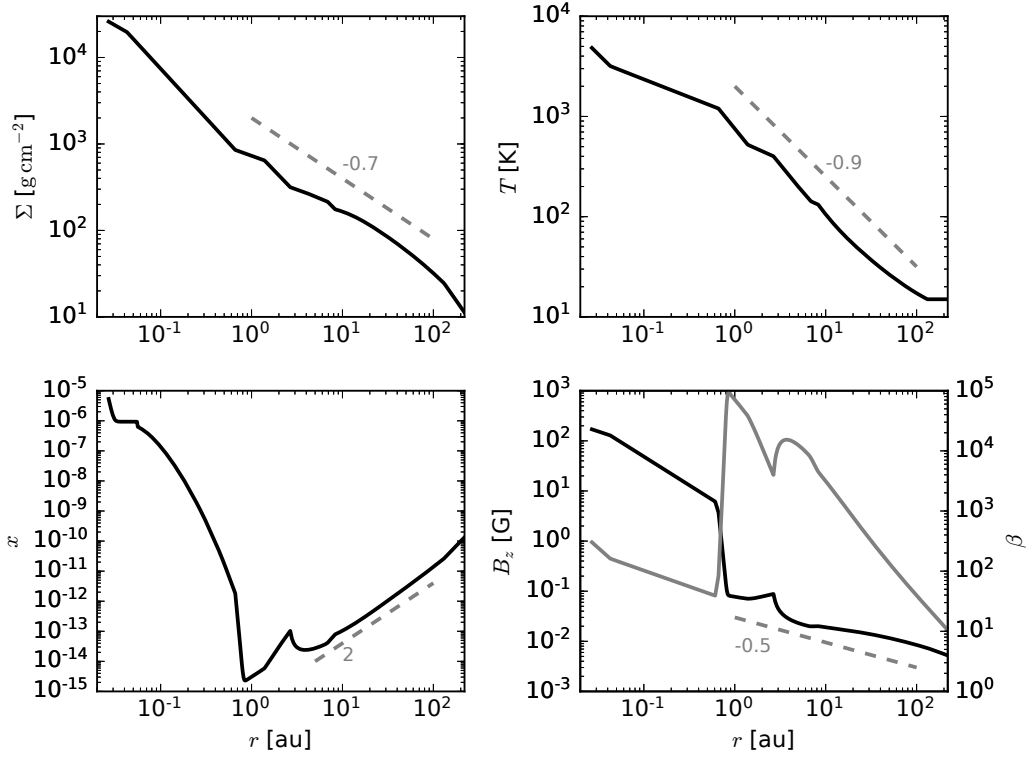}
	\caption{The structure of the disk with $\alpha=0.01$, $\dot{M} = 10^{-7}\,M_{\odot}\,\mathrm{yr}^{-1}$ around a star with $M=1\,M_{\odot}$. Top left: surface density, top right: midplane temperature, bottom left: midplane ionization fraction, bottom right: vertical magnetic field strength (left $y$-axis, black line) and plasma beta (right $y$-axis, grey line). Gray dashed lines with numbers depict typical slopes.}
	\label{Fig0}
\end{figure}

In Figure~\ref{Fig0} we plot the radial profiles of the midplane temperature, surface density, midplane ionization fraction, vertical magnetic field strength and midplane plasma beta for the disk with $\alpha=0.01$, $\dot{M} = 10^{-7}\,M_{\odot}\,\mathrm{yr}^{-1}$. Stellar mass $M=1\,M_{\odot}$. In the simulation, cosmic rays ionization rate $\xi_0=10^{-17}\,\rm{s}^{-1}$ and attenuation length $R_{\rm{CR}}=100\,\rm{g}\,\rm{cm}^{-2}$, stellar X-ray luminosity $L_{\rm{XR}}=10^{30}\,\rm{erg}\,\rm{s}^{-1}$, mean dust grain size $a_{\rm{d}}=0.1\,\mu\rm{m}$.

Figure~\ref{Fig0} shows that the surface density and temperature gradually decrease with distance from the star. The temperature is $\sim 5000$~K near the inner edge of the disk and $15$~K near its outer edge, $r_{\rm{out}}=220$~au. In our model, the radial dependences of all physical quantities are power-law functions of the distance. Indexes of the power laws depend only on the parameters of the opacity. The indexes change throughout the disk as the opacity does. That is why the dependences $T(r)$ and $\Sigma(r)$ appear as piecewise-linear functions in logarithmic scale in Figure~\ref{Fig0}. Typical slope of the temperature profile $p_{\rm{T}}=-0.9$ in range $1-100$~au, and the typical slope of the surface density profile is $p_{\rm{\Sigma}}=-0.7$. The latter is consistent with observations indicating that $p_{\rm{\Sigma}}\in[0.4,\,1]$~\citep{andrews09}.

The radial profile of $B_z$ is more complex. In the innermost par of the disk, $r<0.8$~au, the ionization fraction is high, $x>10^{-10}$ and the magnetic field is frozen in gas. The radial profile of $B_z$ follows the surface density profile in this region, and $B_z\simeq 170$~G at the inner edge of the disk $r_{\rm{in}}=0.027$~au. The ``dead'' zone is situated at $r>0.8$~au, where the ionization fraction is very low. Magnetic ambipolar diffusion reduces the magnetic field strength by 1-2 orders of magnitude in this region, so that typical $B_z(3\,\mathrm{au})=0.1$~G. Near the outer edge of the disk, the magnetic field is frozen in gas and its intensity is $4\times 10^{-3}$~G. The plasma beta is not constant in the disk. It is $\sim 100$ near the inner edge of the disk, $\sim 10^{4}-10^5$ inside the ``dead'' zone and $\sim 10$ near the outer edge of the disk.

\section{Results}
\label{Sec:res}
In this Section, we present the results of simulations of the MFT dynamics in the accretion disks. The system of dynamical equations (\ref{Eq:v}, \ref{Eq:z}, \ref{Eq:rho}, \ref{Eq:T}) is solved with the help of the explicit Runge-Kutta method of the fourth order with automatic selection of the time step and relative accuracy $\varepsilon=10^{-6}$. At each time step, the radius and magnetic field strength of the MFT are calculated with the help of relations (\ref{Eq:a}-\ref{Eq:B}).

We assume that the MFT forms inside the disk at height $z_0=[0.5,\,1,\,1.5]\,H$, in thermal equilibrium with the surrounding gas, $T_0=T_e$, and with velocity $u_0=0$. We specify the initial magnetic field strength of the MFT using plasma beta definition,
\begin{equation}
	\beta_0 = \frac{8\pi P_{\rm{g}0}}{B_0^2},\label{Eq:beta0}
\end{equation}
where $P_{\rm{g}0}$ is the initial gas pressure inside the MFT. Initial density is determined from the condition of the pressure equilibrium (\ref{Eq:pbal}) at $t=0$ in terms of $\beta_0$,
\begin{equation}
\rho_0 = \dfrac{P_{\rm{e}}(z_0)}{\dfrac{R_{\rm{g}}T_0}{\mu}\left(1 + \dfrac{1}{\beta_0}\right)}.
\end{equation}
Adiabatic index of the molecular hydrogen gas $\gamma=7/5$.

\subsection{Fiducial run}
\label{Sec:fiduc}

\begin{figure}[t]
\centering
	\includegraphics[width=14.0cm, angle=0]{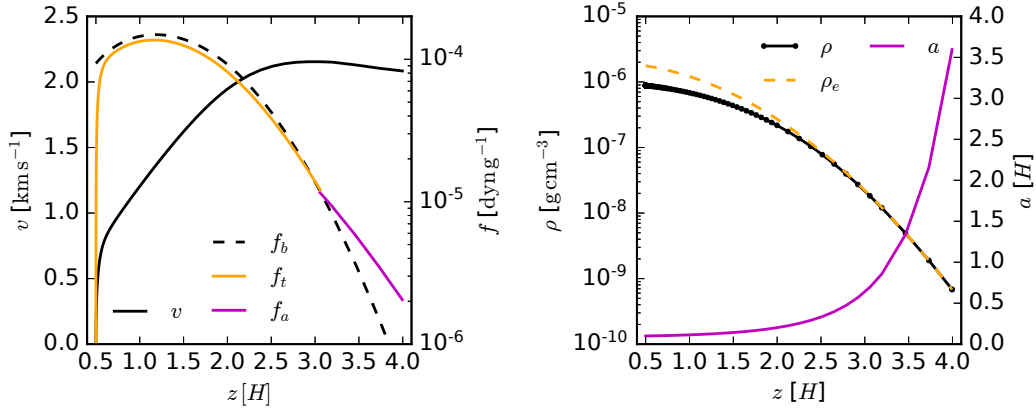}
	\caption{Left panel: vertical profiles of the MFT velocity (left $y$-axis, black line) and forces per unit mass (right $y$-axis, black dashes: buoyancy force, orange: turbulent drag force, magenta: aerodynamic drag force). Right panel: vertical profiles of densities (left $y$-axis, black line with dots~--~MFT density, orange dashes~--~disk density) and MFT radius (right $y$-axis, magenta line). Initial parameters: $a_0=0.1\,H$, $\beta_0=1$, $z_0=0.5\,H$.}
	\label{Fig1}
\end{figure}

Let us first consider the typical picture of MFT dynamics. In the fiducial run, the dynamical equations (\ref{Eq:v}, \ref{Eq:z}, \ref{Eq:rho}, \ref{Eq:T}) are solved assuming that the MFT is at the distance $r=0.027$~au inside the disk. Parameters of the disk at this distance are following: temperature $T_{\rm{e}}=4830$~K, midplane density $\rho_{\rm{m}}=2\times 10^{-6}\,\rm{g}\,\rm{cm}^{-3}$, height scale $H=6.2\times 10^{-4}$~au, magnetic field strength $B_z=170$~G. Initial parameters of the MFT: $a_0=0.1\,H$, $\beta_0=1$, $z_0=0.5\,H$. In Figure~\ref{Fig1}, we plot the profiles of the velocity, drag and buoyancy forces, densities $\rho$ and $\rho_{\rm{e}}$, and MFT radius.

Left panel of Figure~\ref{Fig1} shows that initially MFT accelerates to the velocity $\simeq 0.7\,\rm{km}\,\rm{s}^{-1}$ almost instantly, because the turbulent drag force is much less than the buoyancy force. After this acceleration, the drag force and buoyancy force become nearly equal, $f_t\lesssim f_b$, and the MFT moves with increasing velocity. The acceleration decreases when the MFT rises to height $z\simeq 2.5-3\,H$. Above the disk, $z>3\,H$, the acceleration tends to zero, as the buoyancy and aerodynamic drag forces become very small. The MFT moves by inertia with nearly constant velocity, $v\simeq 2\,\rm{km}\,\rm{s}^{-1}$.

The MFT moves in highly non-uniform medium. Right panel of Figure~\ref{Fig1} shows that the MFT expands during its motion and its density decreases. External density also decreases with height. The densities difference reduces from $50\,\%$ in the starting point to nearly zero at $z>3\,H$, i.e. the degree of the buoyancy decreases. The radius of the MFT becomes larger than the height of the disk at $z\simeq 4\,H$. Further motion of the MFT cannot be investigated in the frame of slender tube approximation. We assume that the MFT dissipates at $z\simeq 4\,H$. The dissipation of the MFTs in the atmosphere of the disk leads to the formation of the non-stationary magnetized corona. The heating of the corona by the dissipation of the magnetic field rising from the disk have also been found and discussed by~\cite{galeev79} and~\cite{stella84} in the context of the accretion disks around black holes and by~\cite{miller00} in application to the disks around classical T~Tauri stars.

\subsection{Dependence on parameters}
\label{Sec:param}

\begin{figure}[t]
\centering
	\includegraphics[width=7.0cm, angle=0]{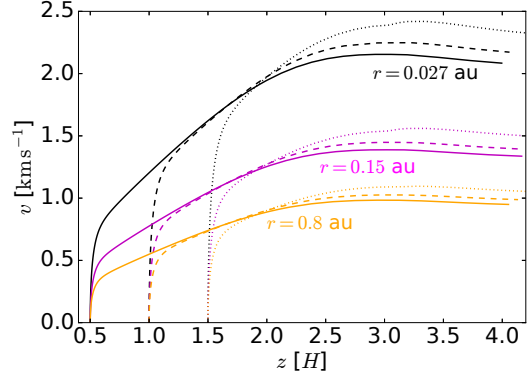}
	\caption{Velocity profiles at the distances $r=0.027$~au (black), $r=0.15$~au (magenta), $r=0.8$~au (orange) for different initial heights (solid lines: $z_0=0.5\,H$, dashes: $z_0=1\,H$, dots: $z_0=1.5\,H$).}
	\label{Fig2}
\end{figure}

Dynamics of MFT depends on the initial position of the MFT inside the disk. In Figure~\ref{Fig2}, we present the dependence of the MFT velocity on the initial height and distance from the star. We consider three distances, $r=0.027$~au (close to the inner boundary of the disk), $r=0.15$~au ($T_{\rm{e}}=2025$~K, $\rho_{\rm{m}}=4.1\times 10^{-8}\,\rm{g}\,\rm{cm}^{-3}$, $H=0.0053$~au, $B_z=29.5$~G) and $r=0.8$~au (outer zone of the thermal ionization region, where $T_{\rm{e}}=970$~K, $\rho_{\rm{m}}=8.3\times 10^{-10}\,\rm{g}\,\rm{cm}^{-3}$, $H=0.045$~au, $B_z=0.14$~G). The dynamics of the MFTs rising from $z_0=0.5\,H$, $z_0=1\,H$ and $z_0=1.5\,H$ is considered.

Figure~\ref{Fig2} shows that the MFTs rapidly accelerate in the beginning, like in the fiducial run (see Section~\ref{Sec:fiduc}). After that, the MFTs rise with increasing velocity. Above the disk, $z>3\,H$, the MFTs move with a constant velocity, that is $\sim 1$~km~s$^{-1}$ at $r=0.8$~au and $2-2.5$~km~s$^{-1}$ at $r=0.027$~au.  Our simulations shows that rise times to the surface of the disk are $0.5\,P_{\rm{orb}}$, $0.8\,P_{\rm{orb}}$ and $1.2\,P_{\rm{orb}}$ for the MFTs with $z_0=0.5\,H$, $z_0=1\,H$ and $z_0=2\,H$, respectively ($P_{\rm{orb}}$ is the rotation period).

\subsection{Terminal velocity of the MFT}
\label{Sec:v_b}
As it was shown in previous Section, after the initial acceleration the MFT moves at a constant velocity, that is determined from the balance between the buoyancy and drag forces. In the case of the aerodynamic drag, we obtain the equality
\begin{equation}
	\Delta\rho g_z = \frac{\rho_{\rm{e}}v_{\rm{b}}^2}{2}\frac{C_{\rm{d}}}{\pi a^2},\label{Eq:fbal}
\end{equation}
where $\Delta\rho = \rho_{\rm{e}}-\rho$. The densities difference in the thermal equilibrium (see Eq. (\ref{Eq:pbal}))
\begin{equation}
	\Delta\rho = \frac{B^2}{8\pi v_{\rm{s}}^2}.\label{Eq:drho}
\end{equation}
Substituting (\ref{Eq:drho}) into Eq. (\ref{Eq:fbal}) it is easy to derive the formula for calculating the terminal velocity~\citep{parker_book}
\begin{equation}
	v_{\rm{b}} = v_{\rm{a}}\left(\frac{\pi}{C_{\rm{d}}}\right)^{1/2}\left(\frac{a}{H}\right)^{1/2}\left(\frac{z_0}{H}\right)^{1/2},\label{Eq:v_b}
\end{equation}
where
\begin{equation}
v_{\rm{a}} = \frac{B}{\sqrt{4\pi\rho}}
\end{equation}
is the Alfv{\'e}n speed. Formula (\ref{Eq:v_b}) shows that the terminal velocity of the MFT with $a\sim 1\,H$ at $z_0\sim 1\,H$ approximately equals $v_{\rm{a}}$. For convenience, we express the terminal velocity in terms of plasma beta and local sound speed $v_{\rm{s}}$
\begin{equation}
	v_{\rm{b}} = v_{\rm{s}}\sqrt{\frac{2}{\beta}}\left(\frac{\pi}{C_{\rm{d}}}\right)^{1/2}\left(\frac{a}{H}\right)^{1/2}\left(\frac{z_0}{H}\right)^{1/2}.\label{Eq:v_b2}
\end{equation}

\begin{figure}[t]
\centering
	\includegraphics[width=7.0cm, angle=0]{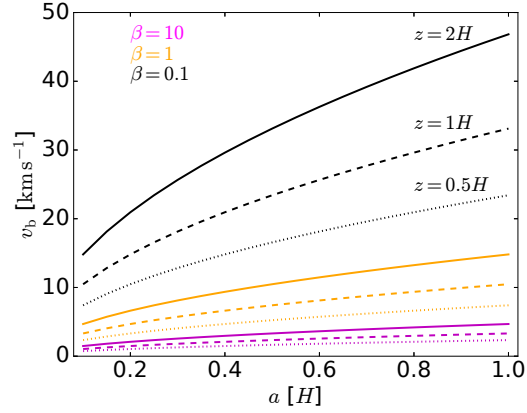}
	\caption{Dependence of the terminal MFT velocity on the radius of the MFT. Black lines: plasma $\beta=0.1$, orange lines: $\beta=1$, magenta lines: $\beta=10$. Solid lines: $z_0=2\,H$, dashed lines: $z_0=1\,H$, dotted lines: $z_0=0.5\,H$. }
	\label{Fig3}
\end{figure}

Figure~\ref{Fig3} shows dependences of the terminal velocity on the MFT radius for various plasma betas and initial heights $z_0$. Terminal velocities range from $2$ to $50$~km~s$^{-1}$ for radii in range $[0.1,\,1]\,H$. The more initial height $z_0$, the more the terminal velocity of the MFT. The MFTs with weak magnetic field ($\beta=10$) move slowly with the velocity smaller than the sound speed. The MFTs with strong magnetic field ($\beta=0.1$) accelerate to the velocities $\sim 40-50$~km~s$^{-1}$ characterizing molecular outflows. Generally speaking, the rising MFT can cause the outflows. 

\subsection{Buoyancy-driven outflows}
\label{Sec:outflows}
In Sections~\ref{Sec:fiduc}-\ref{Sec:v_b} we show that the MFT can form outflows from the disk. The buoyancy extracts the magnetic flux from the disk over the time scale of the MFT rise to the surface of the disk. As we discuss in the introduction, the toroidal magnetic field is permanently amplified in the region of thermal ionization. When the magnetic field reaches state with $\beta\sim 1$, the MFTs form, rise from the disk and carry away the excess of the magnetic flux. Time scale of the magnetic field amplification can be estimated from the induction equation. In the approximations of the accretion disk model, the time scale of the azimuthal magnetic field amplification (see~\cite{fmfadys}),
\begin{equation}
	t_{\rm{gen}} = \frac{2}{3}\frac{|B_{\varphi}|}{|B_z|}\Omega^{-1}\left(\frac{z}{r}\right)^{-1}\simeq 2.12P_{\rm{orb}}\left(\frac{z/r}{0.05}\right)^{-1}\frac{|B_{\varphi}|}{|B_z|}.\label{Eq:t_phi}
\end{equation}
Formula (\ref{Eq:t_phi}) shows that $t_{\rm{gen}}\simeq 2P_{\rm{orb}}$ at height $z=1\,H$.

Comparison of rise times discussed in Section~\ref{Sec:param} and estimate (\ref{Eq:t_phi}) shows that the MFTs rise time is less than $t_{\rm{gen}}$. Therefore, considered buoyancy-driven outflows are periodic. The period of the outflows will be of the order of the magnetic field amplification time scale, i.e. several rotation periods.

We propose that periodically rising MFTs can contribute to the variability of the YSOs radiation. In the region $r=[0.5,\,0.8]$~au, temperature $T\lesssim 1500$~K and the MFTs contain dust grains. Such rising MFTs can absorb stellar radiation and reemit it in infra-red (IR). This process can be responsible for the IR-variability of YSOs. The time scale of the variability would be of order of the magnetic field amplification time scale, i.e. rotation periods. This time scale ranges from several days at $r=0.027$~au to several months at $r=0.8$~au.

\section{Conclusions and discussion}
\label{Sec:discussion}

In this paper, we investigate the dynamics of the magnetic flux tubes in the accretion disks of young stars. In our previous paper, the adiabatic motion of the MFTs was considered~\citep{bmfad}. Now we include in the model the radiative heat exchange between the MFT and surrounding medium. The disk in hydrostatic equilibrium is considered. The density, temperature and magnetic field strength of the disk are calculated with the help of our MHD model of the accretion disks~\citep{fmfadys, kh17}.

We investigate the dynamics of the MFT with various initial radii $a_0$ and plasma beta $\beta_0$ formed at the different distances from the star, $r=[0.027,\,0.15,\,0.8]$~au, and at the different heights above the midplane of the disk, $z_0=[0.5,\,1,\,2]$~$H$. The accretion disk with turbulence parameter $\alpha=0.01$ and accretion rate $\dot{M} = 10^{-7}\,M_{\odot}\,\mathrm{yr}^{-1}$ around solar mass T~Tauri star is considered.

The simulations show that the MFTs rise from the disk to the atmosphere with velocities up to $\simeq 50\,\rm{km}\,\rm{s}^{-1}$. The farther from the star the MFT formed the less its terminal velocity. We divide the MFTs into two categories. Small MFTs (radius less than $\sim 0.1H$) cannot accelerate to speeds more than $10$~km s$^{-1}$. Large MFTs having radii more than $~\sim 0.1\,H$ can reach velocities up to $50$~km~s$^{-1}$. 

The time of rise of the MFT to the surface of the disk is of the order of rotation period. This time is less than the time scale of the toroidal magnetic field amplification, $t_{\rm{gen}}$. Therefore, the MFTs form inside the disk and float from it periodically over time scales $\sim t_{\rm{gen}}$, that ranges from several days to several months in the region of thermal ionization. The MFTs catastrophically expand above the disk.

The MFTs with weak magnetic field ($\beta=10$) rise slowly with speeds less than the sound speed. The MFTs with $\beta=1$ form outflowing magnetized corona. Strongly magnetized MFTs ($\beta=0.1$) cause the outflows with velocities $20-50$~km s$^{-1}$. The outflows velocity is consistent with the velocity of the molecular outflows from YSOs (see reviews~\cite{ray07} and \cite{frank14}).

The MFTs formed in the region of the disk with $T=[1000,\,1500]$~K contain dust particles. The rising MFTs will periodically absorb the stellar radiation and reemit it in IR. We propose that this process can contribute to the observational IR-variability found in many YSOs (see, for example,~\cite{flaherty16}). Shadowing of the outer disk regions by periodically rising MFTs can also appear as the IR-variability of the disk.

It should be noted that the specific mechanism of the magnetic field generation is not important from the point of view of the dynamics of the magnetic flux tubes. Our conclusions can also be generalized for the dynamo generated magnetic field in the disks.

In this work, the isothermal disk was considered. Several works investigated influence of the radiation transfer on the vertical structure of the protoplanetary disks (see, for example,~\cite{andes} and reference therein). We will consider the disk thermal structure in details in our next paper. We will investigate the influence of the magnetic field of the disk on the MFT dynamics. It is also interesting task to compare theoretical variability due to periodically rising MFTs with the IR-periodicity of YSOs.

\normalem
\begin{acknowledgements}
We thank the anonymous referee for some useful comments. This work is supported by the Russian science foundation  (project 15-12-10017). 
\end{acknowledgements}
  
\bibliographystyle{raa}
\bibliography{khaibrakhmanov}

\end{document}